\documentclass[conference]{IEEEtran}
\IEEEoverridecommandlockouts

\usepackage{amsmath,amssymb,amsfonts}
\usepackage{algorithmic}
\usepackage{graphicx}
\usepackage{textcomp}
\usepackage[table,xcdraw]{xcolor}
\def\BibTeX{{\rm B\kern-.05em{\sc i\kern-.025em b}\kern-.08em
    T\kern-.1667em\lower.7ex\hbox{E}\kern-.125emX}}
\usepackage{mathptmx} 
\usepackage{fancyhdr}
\usepackage[normalem]{ulem}
\usepackage[hyphens]{url}
\usepackage[sort,nocompress]{cite}
\usepackage[final]{microtype}
\usepackage{flushend}

\usepackage{adjustbox}
\usepackage{enumitem}
\usepackage{booktabs}
\usepackage{xspace}
\usepackage{multirow}
\usepackage{multicol}
\usepackage{setspace}
 
\usepackage[bookmarks=true,breaklinks=true,letterpaper=true,colorlinks,citecolor=blue,linkcolor=blue,urlcolor=blue]{hyperref}

\begin{document}
\newcommand{\sysname}{MARC\xspace}
\newcommand{\sysnames}{MARC's\xspace}
\makeatletter
\newcommand{\newlineauthors}{%
  \end{@IEEEauthorhalign}\hfill\mbox{}\par
  \mbox{}\hfill\begin{@IEEEauthorhalign}
}
\makeatother


\title{Securing DRA\underline{M} at Scale: \underline{A}RFM-Driven \underline{R}ow Hammer Defense with Unveiling the Threat of Short tR\underline{C} Patterns}

\author
{
\IEEEauthorblockN{\large Nogeun Joo}
\IEEEauthorblockA{\textnormal{nogeun.joo@sk.com}\\
\textnormal{KAIST}\\\textnormal{Daejeon, South Korea}}
\and
\IEEEauthorblockN{\large Donghyuk Kim}
\IEEEauthorblockA{\textnormal{kar02040@kaist.ac.kr}\\
\textnormal{KAIST}\\\textnormal{Daejeon, South Korea}}
\and
\IEEEauthorblockN{\large Hyunjun Cho}
\IEEEauthorblockA{\textnormal{h.cho@kaist.ac.kr}\\
\textnormal{KAIST}\\\textnormal{Daejeon, South Korea}}
\newlineauthors
\IEEEauthorblockN{\large Junseok Noh}
\IEEEauthorblockA{\textnormal{junseok1.noh@sk.com}\\
\textnormal{SK hynix Inc.}\\\textnormal{Icheon, South Korea}}
\and
\IEEEauthorblockN{\large Dongha Jung}
\IEEEauthorblockA{\textnormal{dongha1.jung@sk.com}\\
\textnormal{SK hynix Inc.}\\\textnormal{Icheon, South Korea}}
\and
\IEEEauthorblockN{\large Joo-Young Kim}
\IEEEauthorblockA{\textnormal{jooyoung1203@kaist.ac.kr}\\
\textnormal{KAIST}\\\textnormal{Daejeon, South Korea}}
}

\maketitle


\begin{abstract}
Since the disclosure of the row hammer (RH) attack phenomenon in 2014, a significant threat to system security, it has been active research in both industry and academia. The major issue with RH is the deterioration of cell endurance as DRAM technology advances, making the defense system more challenging due to the reduced threshold. With the growing challenges posed by malicious RH strategies, various RH mitigation IPs have been proposed. Meanwhile, a new DRAM feature known as refresh management (RFM) was incorporated into the JEDEC standards for DDR5/LPDDR5, which allows the memory controller to have the necessary time to respond to RH. However, indiscriminate use of RFM leads to reduced system performance and higher power consumption. To address the issue of powerful RH attacks, our study involved an extensive analysis of the prevalent attack patterns in the field. We discovered a strong correlation between the timing and density of the active-to-active command period, ${tRC}$, and the likelihood of RH attacks. Leveraging this correlation, we developed a method to optimize the use of adaptive refresh management (ARFM), thereby maximizing its efficacy. 

In this paper, we introduce \sysname, an innovative ARFM-driven RH mitigation IP that significantly reinforces existing RH mitigation IPs. \sysname dynamically adjusts the frequency of RFM in response to the severity of the RH attack environment, offering a tailored security solution that not only detects the threats but also adapts to varying threat levels. \sysnames detection mechanism has demonstrated remarkable efficiency, identifying over 99\% of attack patterns. Moreover, \sysname is designed as a compact hardware module, facilitating tight integration either on the memory controller-side or DRAM-side within the memory system. It only occupies a negligible hardware area of 3363~\textit{$\mu m^2$}. By activating ARFM based on \sysnames detection, the additional energy overhead is also negligible in normal workloads. We conduct experiments to compare the highest row count throughout the patterns, defined as max exposure, between the vanilla RH mitigation IPs and the \sysname-enhanced versions of the same IPs, focusing on both DRAM-side and memory controller-side. On the DRAM-side, \sysname + probabilistic scheme and \sysname + counter-based tracking scheme achieve 8.1$\times$ and 1.5$\times$ improvement in max exposure ratio compared to the vanilla IPs, respectively. On the memory controller-side, the \sysname + PARA and \sysname + Graphene achieve 50$\times$ and 5.7$\times$ improvement in max exposure ratio compared to the vanilla IPs, respectively. \sysname ensures optimal security without sacrificing system performance, making \sysname a pioneering solution in the realm of RH attack mitigation. 

\end{abstract}


\section{Introduction}

Dynamic random access memory (DRAM), which serves as a primary memory in computer systems, has progressively increased its density with scaling to a smaller form factor over decades~\cite{chang2017understanding, chang2016understanding}. While this trend in DRAM allows for a higher capacity and speed due to its ability to hold a larger amount of memory in a smaller space, it has unexpectedly suffered from reliability issues. First, the reduced size of a cell can contain only a small amount of electrical charge, diminishing its ability to resist noise, thereby increasing the risk of data corruption~\cite{cha2011dram, mandelman2002challenges, yoon2013flash}. Second, the shortened distance between cells can lead to unexpected electromagnetic interference due to coupling effects~\cite{cha2011dram, mandelman2002challenges, konishi1989analysis, redeker2002investigation}. Third, the pursuit of smaller cell dimensions in manufacturing processes leads to an increased presence of cells that are excessively vulnerable to the effects of inter-cell cross-talk, thereby exacerbating the issues. Due to these reasons, contemporary DRAM technologies face a novel issue known as row hammer (RH)~\cite{kim2014flipping, mutlu2017rowhammer,orosa2021deeper}. The RH refers to a phenomenon where repetitive activation of specific cells results in the loss of information in adjacent cells. With the trend of scaling DRAM smaller, the RH phenomenon occurs more readily, and it has been exploited by malicious attackers as part of hardware attack techniques, posing a critical threat to any computer system, including personal computers, edge devices, and cloud servers. 

\begin{table}[b]
\centering
\vspace{-0.1in}
\caption{Previous Row Hammer Mitigation IPs}
\vspace{-0.1in}
\includegraphics[width=\columnwidth]{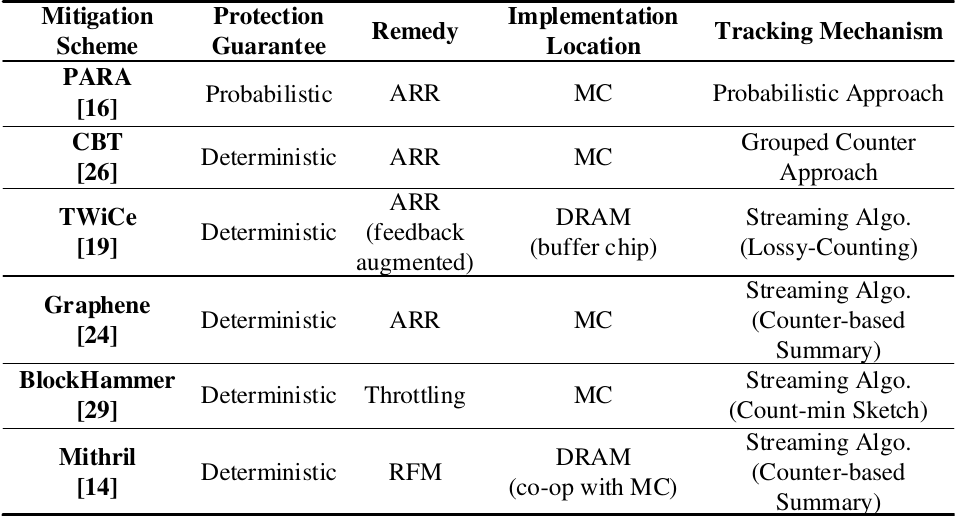}
\vspace{-0.2in}
\label{table_RHIPs}
\end{table}

Various studies have been conducted to resolve the RH~\cite{kim2014flipping, mutlu2017rowhammer,aweke2016anvil,orosa2021deeper,kim2014architectural,you2019mrloc,park2020graphene,kim2022mithril,lee2019twice,yauglikcci2021blockhammer,son2017making,seyedzadeh2016counter}. Among them, architecture-based solutions, particularly favored for their efficiency, can be divided into two main categories: probabilistic~\cite{kim2014flipping,you2019mrloc} and counter-based tracking schemes~\cite{park2020graphene, kim2022mithril, lee2019twice,yauglikcci2021blockhammer,son2017making,seyedzadeh2016counter}. The major RH mitigation IPs studied so far are summarized in Table~\ref{table_RHIPs}. The probabilistic scheme involves refreshing adjacent rows at a certain probability upon receiving an activation command. This method has gained attention for its low area overhead and efficiency. Alternatively, the counter-based tracking scheme counts how often each row is activated and refreshes the row when it reaches a certain threshold. It provides high reliability and protects DRAM from malicious RH attacks~\cite{frigo2020trrespass,aga2017good,gruss2018another,jattke2022blacksmith,van2018guardion,cojocar2019exploiting,gruss2016rowhammer}. Numerous studies to date have aimed to block the attacks with high precision and low cost.

\begin{figure}[!t]
\centering
\includegraphics[width=0.9\columnwidth]{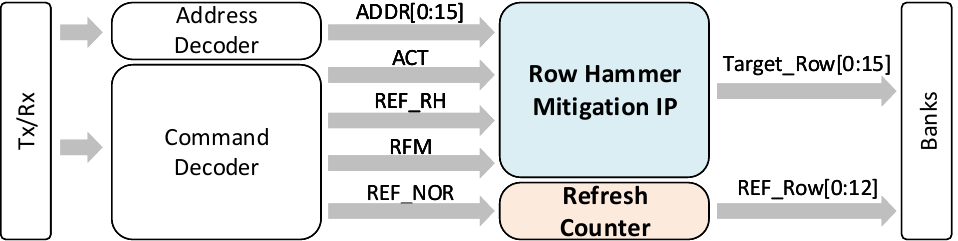}
\caption{DRAM Refresh Operation}
\vspace{-0.25in}
\label{figure_dramrefreshoperation}

\end{figure}

While prior research confirms the effectiveness of RH mitigation, significant drawbacks in counter-based tracking and probabilistic schemes persist. The counter-based method, though effective, incurs substantial area overhead due to extensive row tracking tables, a need intensified by the decreasing cell endurance in modern DRAM. This reduction in endurance necessitates larger tables to manage more rows, enlarging hardware size even under standard conditions. 
Conversely, the probabilistic approach, despite low hardware overhead, suffers from high power consumption due to frequent refresh commands needed to combat cell endurance decline. Both strategies overprotect, especially in typical workloads, preparing for worst-case RH attacks. This results in unnecessary defensive operations and overhead, despite the feasibility of smaller tables or less frequent refreshes under normal conditions. These limitations highlight the challenges of current RH mitigation strategies as cell endurance continues to decline.

To this end, we propose \sysname, a novel adaptive refresh management (ARFM) -driven RH mitigation IP, advancing from the conventional hardware-centric security approach~\cite{jedeclpddr5, jedecddr5}. \sysname accurately distinguishes normal workloads from malicious RH attack patterns by selectively harnessing system-level support. It effectively eliminates concerns such as power consumption and performance degradation associated with ARFM during normal workloads. Moreover, \sysnames main purpose is to augment and strengthen existing RH mitigation IPs, which have inherent limitations despite ongoing development. It is readily integratable with any RH mitigation IP. The integration of RH mitigation IP with \sysname does not necessitate separate port alignment or the use of interfaces. The enhancement of RH mitigation IP's defensive efficacy is achievable solely through its combination with \sysname.

We integrate \sysname with the representative IPs, the counter-based tracking scheme Graphene~\cite{park2020graphene}, and the probabilistic scheme PARA~\cite{kim2014flipping}. We demonstrate improved performances when these IPs leverage ARFM from the memory controller (MC). This approach significantly alleviates the burden on hardware in terms of performance and area while establishing a robust defense system against RH attacks, surpassing hardware-only defense mechanisms. We summarize the key contributions of \sysname below.
\begin{itemize}
    \item Low-power ARFM support: \sysname does not generate additional refresh commands in normal workloads by exploiting ARFM. The power consumption is extremely low even in RH attack pattern.
    \item Low area overhead: \sysname is 23\% smaller than the state-of-the-art counter-based tracking scheme~\cite{park2020graphene}, and incurs no additional overhead in the DRAM's scaling issue.
    \item Portability: \sysname can be integrated with existing RH mitigation IPs complementarily, reducing max exposure ratio up to 50$\times$.
\end{itemize}

\section{Background}
\label{sec_background}


\begin{figure}[!t]
\centering
\includegraphics[width=0.9\columnwidth]{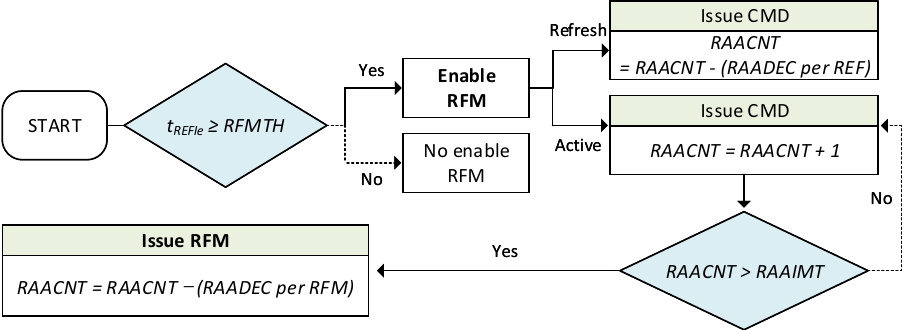}
\vspace{-0.15in}
\caption{RFM Operation Flow}
\vspace{-0.2in}
\label{figure_RFMoperationflow}
\end{figure}

\subsection{DRAM's New Feature: RFM}
\label{subsec_RFM}
In response to the emergence of malicious RH attacks that threaten system security, MC and DRAM have included RFM as a new feature in the JEDEC standard for DDR5/LPDDR5~\cite{marazzi2022protrr,kim2022mithril,kim20231,jedeclpddr5,jedecddr5}. Prior to the introduction of RFM, the DRAM would perform normal refresh operations to secure cell retention time during the refresh operation time $t_{RFC}$ while simultaneously conducting neighboring row refresh (NRR) to cure detected victims through time division, as illustrated in Figure~\ref{figure_dramrefreshoperation}~\cite{kim20231}. Unfortunately, this process required waiting for a time duration equivalent to or exceeding the minimum refresh interval $t_{REFi}$ to perform NRR. To maximize the opportunity for executing NRR and enhance system security, MC and DRAM have included RFM in the JEDEC standard, positioning RFM as a potent tool for bolstering system security against malicious RH attacks in the near future.

\textbf{RFM Operation.}
The operation flow of RFM is depicted in Figure~\ref{figure_RFMoperationflow}. For RFM to be enabled, $t_{REFi}$ must be smaller than RFM threshold (RFMTH). RFMTH is defined as the rolling accumulated active initial management threshold (RAAIMT) $\times$ $t_{RCmin}$. When RFM is enabled, the MC counts the rolling accumulated active count (RAACNT) value, which is the number of active (ACT) commands applied to each bank. RAACNT is then compared with the RAAIMT value, which is the minimum number of ACT command thresholds stored in the DRAM. If the RAACNT value surpasses the RAAIMT, the RFM Command is prepared. The RFM command is issued to the DRAM within a range smaller than the rolling accumulated active maximum management threshold (RAAMMT), considering the controller's schedule. This allows the DRAM, in addition to the traditional method of curing the victim row through time-division with the refresh (REF) command, to gain an additional opportunity to cure the victim row with the RFM command. The RAACNT is managed in a decreasing manner upon the issuing of the REF and RFM commands. The value used for decrementing RAACNT is known as RAA decrement (RAADEC). As depicted in Figure~\ref{figure_RFMoperationflow}, for LPDDR5, RAADEC is configured with two distinct values: one for the REF command and another for the RFM command.

\textbf{DRFM.}
   Directed RFM (DRFM) has been proposed as a means of delivering detected RH aggressors from the RH defense scheme installed on the MC-side to the DRAM. When DRFM is enabled, the detected RH aggressors are delivered to the DRAM along with the RFM command, as shown in Figure~\ref{figure_ADRFM} (a). To enable DRFM, it is essential that RH mitigation IP  
  specific to the MC-side is installed, separate from the RH mitigation IP on the DRAM-side.
  
\begin{figure}[!t]
\centering
\includegraphics[width=\columnwidth]{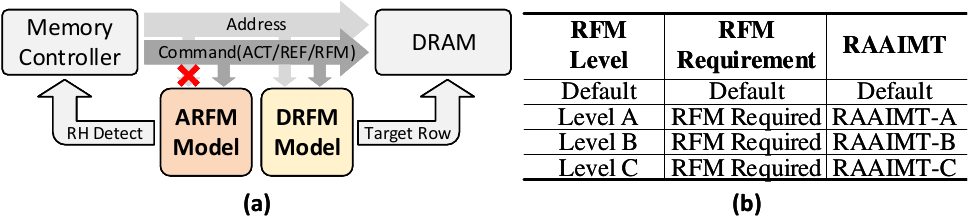}
\vspace{-0.3in}
\caption{(a) ARFM and DRFM (b) ARFM Level}
\vspace{-0.2in}
\label{figure_ADRFM}
\end{figure}

\textbf{ARFM.} The MC can adaptively change how often it sends RFM commands using ARFM. The MC can change the value of RAAIMT depending on the situation. As shown in Figure~\ref{figure_ADRFM} (b), the mode register of DRAM stores not only RAAIMT, which defines the basic operation of RFM, but also the values of RAAIMT-A, -B, and -C according to the level of ARFM. Level-C is the highest level and requires the highest number of RFMs issued. ARFM can only operate when RFM is enabled. ARFM does not necessarily require RH mitigation IP to be installed inside the MC, and therefore, there is no need to deliver the aggressor to the DRAM, as shown in Figure~\ref{figure_ADRFM} (a). By adjusting the RAAIMT through the levels in the mode register, more curing opportunities can be provided inside the DRAM. From the perspective of the DRAM, this results in the benefit of increased refresh effects.

\subsection{Key Parameters Related to Row Hammer}
Figure~\ref{figure_keyparam} illustrates the main timing parameters of DRAM related to RH attacks. An ACT command can be reapplied after satisfying the row active time $t_{RAS}$ and row precharge time $t_{RP}$, meaning that $t_{RAS}$ + $t_{RP}$ forms the active command period, which is denoted as $t_{RC}$. In normal workloads, various functions including row access and basic DRAM operations such as read/write are performed, necessitating different settings for the $t_{RC}$ value. However, since the sole purpose of an RH attack is to access rows, a shorter $t_{RC}$ time is a key to allow more row accesses within the fixed time $t_{REFW}$. Therefore, RH attacks typically utilize short $t_{RC}$ values close to the specified minimum $t_{RC}$. Additionally, a longer $t_{REFW}$ makes the system more vulnerable to RH attacks by allowing more ACT commands.

\begin{figure}[!t]
\centering
\includegraphics[width=\columnwidth]{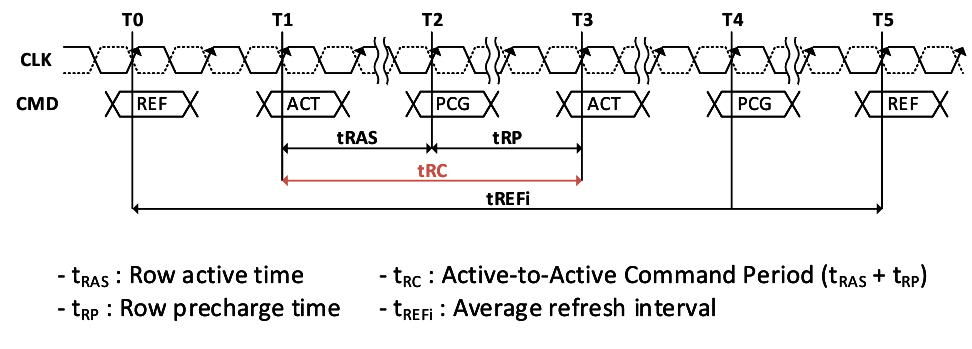}
\vspace{-0.3in}
\caption{Key Timing Parameters of DRAM}
\vspace{-0.2in}
\label{figure_keyparam}
\end{figure}





\subsection{Max Exposure}
We introduce a new metric known as \textbf{max exposure} to evaluate the effectiveness of RH defenses. As shown in Figure~\ref{figure_Maxexposure}, this metric measures the number of ACT commands for each row within a memory pattern, with the count for a specific row being reset when it is refreshed by the RH mitigation IP during the refresh window $t_{REFW}$. It involves identifying the row with the highest count throughout the duration of the pattern run within $t_{REFW}$ and recording this peak value. A lower max exposure score indicates a stronger RH defense performance. 


We evaluated the defensive efficacy of RH mitigation systems using the maximum exposure ratio (MER) for relative comparison. This metric, expressed as a ratio, helps us assess the impact of \sysname on system resilience against RH attacks. The MER establishes the max exposure of the RH mitigation IP with $t_{RC}$ set to 60$ns$ and 50 aggressors, normalized to a baseline value of 1. Subsequent max exposures under varying conditions are normalized against this baseline. This approach allows for precise comparison of RH defense effectiveness across different scenarios by scaling exposure levels to a uniform reference point.

 
\section{Motivation}
\subsection{Limits of Counter-based and Probabilistic Defense Mechanisms}


To counter RH attacks, defenses typically use hardware-level mitigation IPs and system-level refresh period modulation. However, accelerating the refresh cycle at the system level significantly impacts power consumption. Thus, recent research focuses on enhancing hardware-based defenses without system-level support.

\begin{figure}[!t]
\centering
\includegraphics[width=\columnwidth]{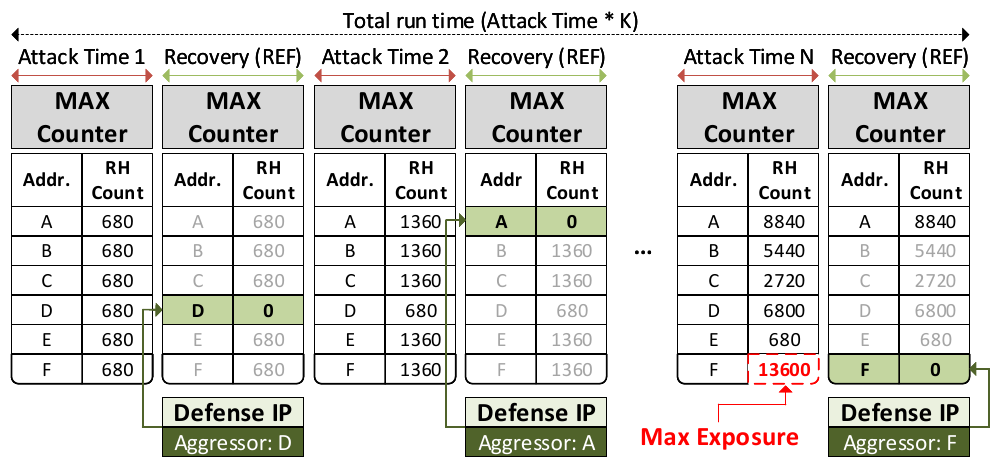}
\vspace{-0.25in}
\caption{Max Exposure}
\vspace{-0.2in}
\label{figure_Maxexposure}
\end{figure}

As DRAM technology continues to advance, it consistently reveals a decrease in cell endurance, suggesting a limit to the effectiveness of hardware-based defense improvements. We have illustrated the challenges posed by reduced cell endurance on previous defense mechanisms in Figure~\ref{figure_motivation1}. Firstly, the counter-based tracking scheme must deal with the need to increase the size of the tracking table as cell endurance diminishes. This issue becomes particularly critical as we aim for higher speeds and densities in DRAM, which significantly increases the area overhead and consequently raises costs. Figure~\ref{figure_motivation1} (a) demonstrates that to accommodate the decreased cell endurance, the counter-based tracking scheme must enlarge its table size to protect against all possible RH aggressors. The graph indicates an expected 32$\times$ increase in table size when cell endurance is at 1.5K, compared to a cell endurance of 50K.


\begin{figure}[!b]
\centering
\vspace{-0.15in}
\includegraphics[width=\columnwidth]{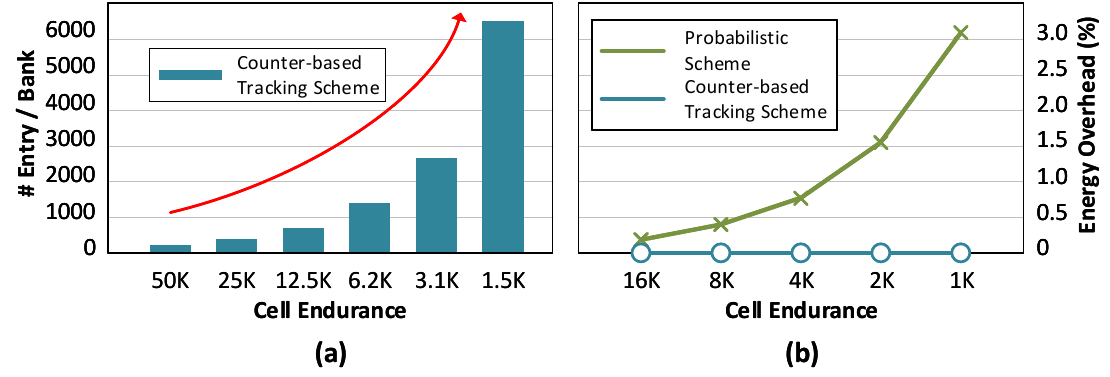}
\caption{Overhead of RH mitigation IPs on Different Cell Endurance. (a) Energy Overhead of Probabilistic vs. Counter-based Tracking (b) Table Size Overhead of Counter-based Tracking}
\label{figure_motivation1}
\end{figure}

Second, the probabilistic scheme necessitates a rise in the frequency of REF commands to improve random sampling effectiveness. Although this scheme poses a smaller area overhead than counter-based tracking, it paradoxically trends against the industry's drive for low-power consumption devices. Figure~\ref{figure_motivation1} (b) presents an analysis of the energy overhead under normal workloads for both IPs, configured to constrain max exposure within the limits of cell endurance during RH attack profiles. The counter-based tracking scheme, as exemplified by Graphene~\cite{park2020graphene}, operates negligibly under normal workloads, thereby accruing no significant energy overhead independent of cell endurance. Conversely, the probabilistic scheme, as exemplified by PARA~\cite{kim2014flipping}, operates at a constant, regardless of workload, resulting in an energy overhead that spikes as cell endurance decreases. Our data indicates that when cell endurance decreases from 16K to 1K, there is an observed escalation in energy overhead by more than 16$\times$ under normal workloads.

\subsection{Wasting Resources on Normal Workload}

Despite the infrequency of malicious RH attacks, hardware-based defenses often seem overly robust, tailored more for extreme scenarios than everyday operations. Such overengineering is evident when evaluating RH mitigation IPs under normal workload conditions. Figure~\ref{figure_ME_normal} shows that under normal workload conditions, Graphene's table size, designed for worst-case scenarios, can be reduced by an eighth without compromising security. Similarly, the maximum exposure for PARA, with minimal area overhead, closely matches that of Graphene. Although counter-based schemes like Graphene are traditionally viewed as more effective against RH attacks, this advantage diminishes outside-targeted attacks. This highlights the need for a shift in focus from exclusively strengthening hardware to exploring innovative RH mitigation strategies.

\begin{figure}[!t]
\centering
\includegraphics[width=\columnwidth]{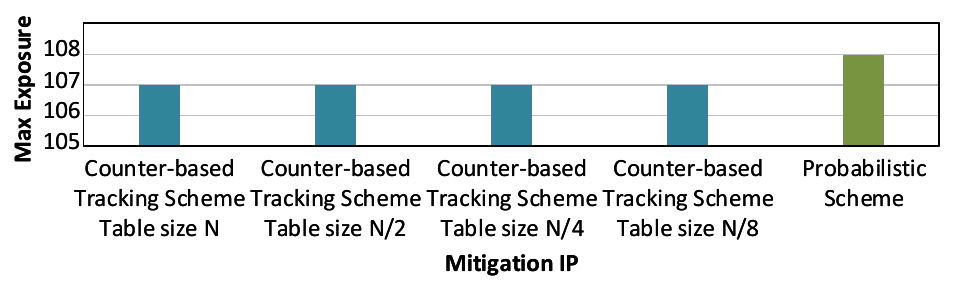}
\vspace{-0.25in}
\caption{Max Exposure in Normal Workload of RH mitigation IPs}
\vspace{-0.15in}
\label{figure_ME_normal}
\end{figure}

\subsection{Transition to ARFM-driven System-supported Protection}
Maintaining memory and system security against RH attacks requires ongoing hardware improvements. However, hardware schemes face limitations like size and energy overheads, impacting system costs. Future RH defenses must reduce these burdens with system-level support.

The JEDEC standard RFM command in DDR5/LPDDR5 devices provides DRAM with system-level assistance, offering dedicated time for RH defense similar to an increased refresh rate. However, RFM must be used cautiously due to potential power and performance impacts.

Given robust existing hardware, the need for RFM under normal workloads is low. Yet, for defending against malicious RH attacks, proactive system-level support via RFM is essential. ARFM meets these conditions, detecting malicious RH attacks and leveraging MC for RH defense while minimizing RFM's side effects during normal workloads.

\section{Observation}
\label{observation_sec}

Our study introduces a new method for enhancing defenses against RH attacks. We performed detailed analyses on normal and RH attack patterns across two device platforms, using actual measuring equipment, and assessed the improvements using our in-house tool. Given the use of DRAM companies' exclusive equipment and experiments simulating RH mitigation IP in DRAM, we must keep the exact data confidential to prevent potential attackers. Hence, for security reasons, we present most data as ratios to a reference value, not revealing absolute numbers.

\subsection{A New Distinguishing Factor for Malicious RH Attack Patterns: $t_{RC}$}
\label{subsec_Obs_a}

Previous research on RH~\cite{kim2014flipping, mutlu2017rowhammer,aweke2016anvil,orosa2021deeper,kim2014architectural,you2019mrloc,park2020graphene,kim2022mithril,lee2019twice,yauglikcci2021blockhammer,son2017making,seyedzadeh2016counter} has primarily focused on either how RH mitigation IPs can effectively detect aggressors or the study of how to evade these detection mechanisms. It has been believed that the key components of the RH attack pattern consist of ACT commands and addresses, with REF commands triggered in synchronization with $t_{REFi}$. Although both attackers and defenders have concentrated their research on aggressors, defining the malicious RH attack pattern solely through aggressors is no longer appropriate. This is because as cell endurance decreases, the number of aggressors will continue to increase, and their driving mechanisms can vary widely. Therefore, we aim to define the RH attack pattern through a different approach.

\begin{figure}[!t]
\centering
\includegraphics[width=\columnwidth]{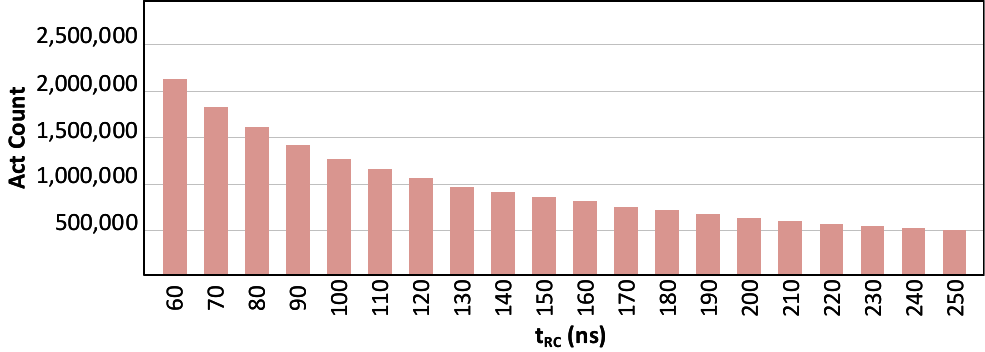}
\vspace{-0.3in}
\caption{Number of ACT Commands on Various $t_{RC}$}
\vspace{-0.2in}
\label{figure_tRCPeriod}
\end{figure}

Figure~\ref{figure_tRCPeriod} illustrates the number of ACT commands issued during $t_{REFW}$ based on $t_{RC}$ values. This number is closely related to the level of RH attacks. From the defender's perspective, attacks based on $t_{RCmin}$ (60${ns}$ for LPDDR5) will be the worst case, and as $t_{RC}$ increases, the RH attack level becomes less severe. To verify whether this assumption corresponds to the RH attack patterns, we analyzed representative RH attack patterns that sufficiently represent diversity using the DRAM maker's actual mobile workload analysis equipment.

We conducted a workload analysis of the most potent RH attack pattern disclosed to date. This pattern, though not encompassing all RH methodologies, caused the failure of a major DRAM manufacturer's chip, making it uniquely impactful. We focused on this pattern to explore mitigation strategies. Given the evolving nature of attack patterns, it is crucial to continuously update and reinforce defense systems to counter these threats effectively.

We sampled a portion of the RH attack pattern (approximately 3${ms}$), extracted all $t_{RC}$ values, and graphically represented the results in Figure~\ref{figure_tRCRHPattern}. About 99\% or more of the $t_{RC}$ values approximate $t_{RCmin}$, and with very fine tolerances, they are consistently issued at a certain value. This also indicates that to maintain short $t_{RC}$, the occupancy rate of ACT commands in the entire pattern is very high. In summary, malicious RH attack patterns consist of short $t_{RC}$ approximating $t_{RCmin}$ and exhibit a high occupancy rate of ACT commands in the overall pattern. To further substantiate this observation, we measured the $t_{RC}$ distribution of normal workload in Section~\ref{subsec_trc_norm} using LPDDR5-based systems, which have sufficient representativeness where the data is based on actual measurements.

\begin{figure}[!b]
\centering
\includegraphics[width=\columnwidth]{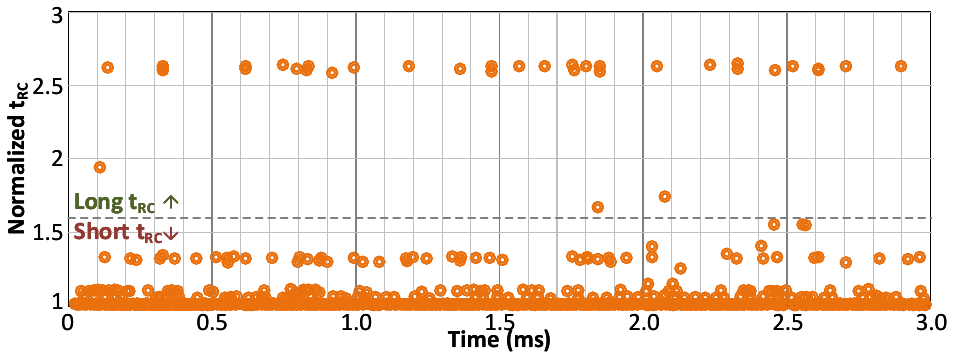}
\caption{$t_{RC}$ Distribution of Row Hammer Attack Pattern}
\label{figure_tRCRHPattern}
\end{figure}

\subsection{Short $t_{RC}$ Distribution in Normal Workload}
\label{subsec_trc_norm}

We selected two device platforms and measured $t_{RC}$ values for various mobile workloads using real equipment. We analyzed short $t_{RC}$ behavior in normal workloads over approximately four-second scenarios, which is sufficient for effective assessment. Figure~\ref{figure_tRCNWPattern} and ~\ref{figure_shortTrc} show the proportion of short $t_{RC}$ in two different normal workloads. The results may vary depending on the equipment. Figure~\ref{figure_tRCNWPattern} counts short $t_{RC}$ (< 150${ns}$) per interval for 20 workloads on the first platform. Figure~\ref{figure_shortTrc} shows the ratio of short $t_{RC}$ (< 100${ns}$) over time for 26 workloads on the second platform. Both platforms show a very low proportion of short $t_{RC}$, less than 1\%. The difference in $t_{RC}$ patterns between normal workloads and malicious RH attacks, which have predominantly short $t_{RC}$, validates using $t_{RC}$ to distinguish between them.

\subsection{Dependency Between $t_{RC}$ and Max Exposure}

In Sections~\ref{subsec_Obs_a} and~\ref{subsec_trc_norm}, we established that short $t_{RC}$ is a significant factor in RH attack patterns. Furthermore, we discover the two critical dependencies between $t_{RC}$ and max exposure, which is the main target point of our proposed solution. First, $t_{RCmin}$ represents the most vulnerable condition for malicious RH attacks, irrespective of the mitigation mechanism used. Second, when $t_{RC}$ exceeds a certain level, the RH attack potency drops to a manageable level that RH mitigation IPs with less extreme designs can handle.

In order to substantiate our findings, we conduct a set of experiments to scrutinize how RH attack levels vary with $t_{RC}$ values from DRAM's perspective, thereby understanding the detrimental effects of short $t_{RC}$ in greater detail. For the memory-side simulation modeling, we apply two factors from the previous research, indicating that 1) victim addresses are alleviated through time division during $t_{RFC}$ when a REF command is applied, and 2) the specific implementation varies but generally adopts either a probabilistic or a counter-based tracking scheme as the defense mechanism. Therefore, we simulated DRAM's RH mitigation IP by implementing two straightforward schemes: a probabilistic scheme and a counter-based tracking scheme. We created a DRAM simulation model configured to perform NRR once every ten refreshes. The probabilistic scheme was set to select random times within 10 $\times$ $t_{REFi}$ and sample the closest authorized ACT address, while the counter-based tracking scheme was designed with a table size of 214. We conducted MER evaluations with 50 multi-sided attack patterns for the wide range of $t_{RC}$ period, with the entire duration of RH attacks being 512${ms}$ (${tREFW}$) and $t_{REFi}$ set to 15.6${us}$.

\begin{figure}[!t]
\centering
\includegraphics[width=\columnwidth]{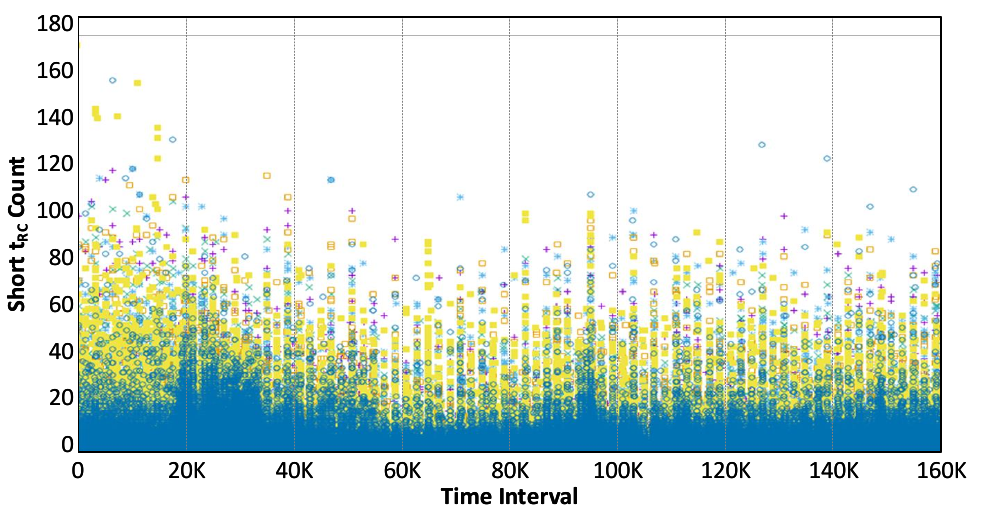}
\vspace{-0.2in}
\caption{Short $t_{RC}$ Distribution of Normal Mobile Workload}
\vspace{-0.18in}
\label{figure_tRCNWPattern}
\end{figure}

\begin{figure}[!t]
\centering
\includegraphics[width=\columnwidth]{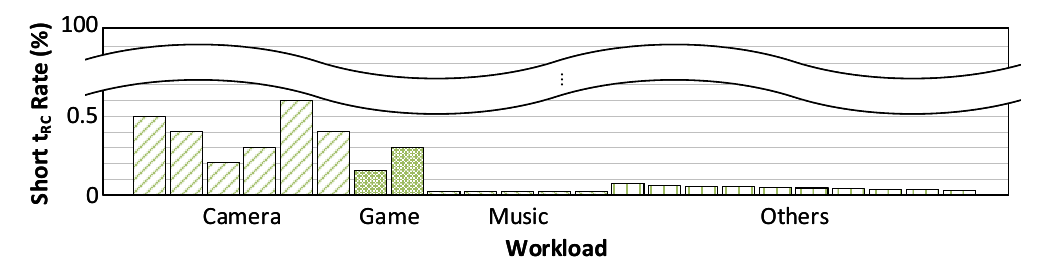}
\vspace{-0.33in}
\caption{Short $t_{RC}$ Analysis on Normal Mobile Workload}
\label{figure_shortTrc}
\end{figure}

Figure~\ref{fig_MER_NRR_para_gra} displays the MER results based on the $t_{RC}$ value on the DRAM side. The probabilistic scheme performs NRR at 10$t_{REFi}$, regardless of the attack pattern. Thus, as $t_{RC}$ increases, the probability of random sampling increases since fewer aggressors can fit within the 10 $t_{REFi}$ window. This results in a significant decrease in MER, as shown in Figure~\ref{fig_MER_NRR_para_gra} (a). The counter-based tracking scheme also shows the highest MER at $t_{RCmin}$ but does not exhibit the dramatic decrease seen in the probabilistic scheme. This is because the counter-based tracking scheme has its own logic threshold, and all aggressors are managed based on that threshold. However, as the $t_{RC}$ value increases, the speed of reaching the logic threshold slows down, which can be seen in the graph where the number of NRR executions is inversely proportional to the increase in $t_{RC}$, as shown in Figure~\ref{fig_MER_NRR_para_gra} (b). Despite minor variances due to the distinct characteristics of RH mitigation mechanisms, both approaches exhibit the highest MER at $t_{RCmin}$. This finding implies that RH attack patterns composed of $t_{RCmin}$ represent the most vulnerable condition for malicious RH attacks, irrespective of the mitigation mechanism used. Conversely, when $t_{RC}$ exceeds 100${ns}$, the RH attack potency drops to 60\% according to the probabilistic scheme, and it is observed that the attack strength reduces to a manageable level within the logic threshold as per the counter-based tracking scheme. The graph we present is based on 50 multi-side attack patterns, but similar trends were observed when assessing variations in the aggressors, consistent with the patterns shown in Figure~\ref{fig_MER_NRR_para_gra}. Consequently, we empirically set the short $t_{RC}$ range to 100${ns}$ or less.

\begin{figure}[!b]
\centering
\includegraphics[width=\columnwidth]{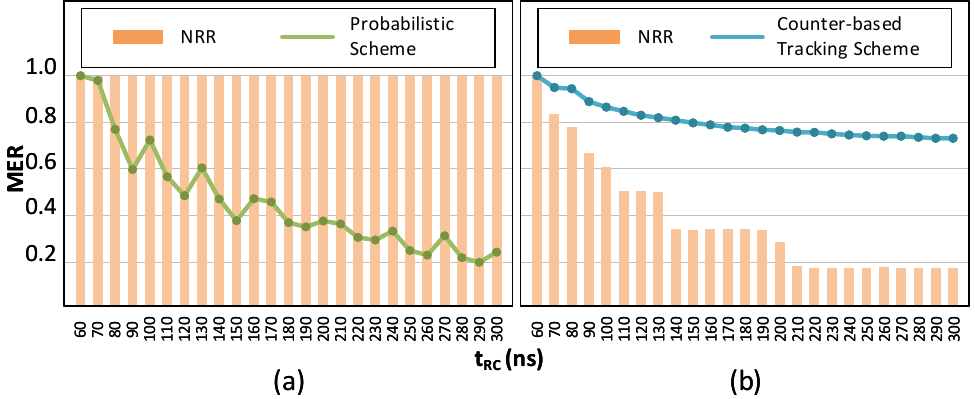}
\caption{Max Exposure Ratio under Neighboring Row Refresh (a) Probabilistic (b) Counter-based Tracking}
\label{fig_MER_NRR_para_gra}
\end{figure} 
\section{\sysname Architecture}
\label{sec_arch}
\begin{figure}[!t]
\centering
\includegraphics[width=0.9\columnwidth]{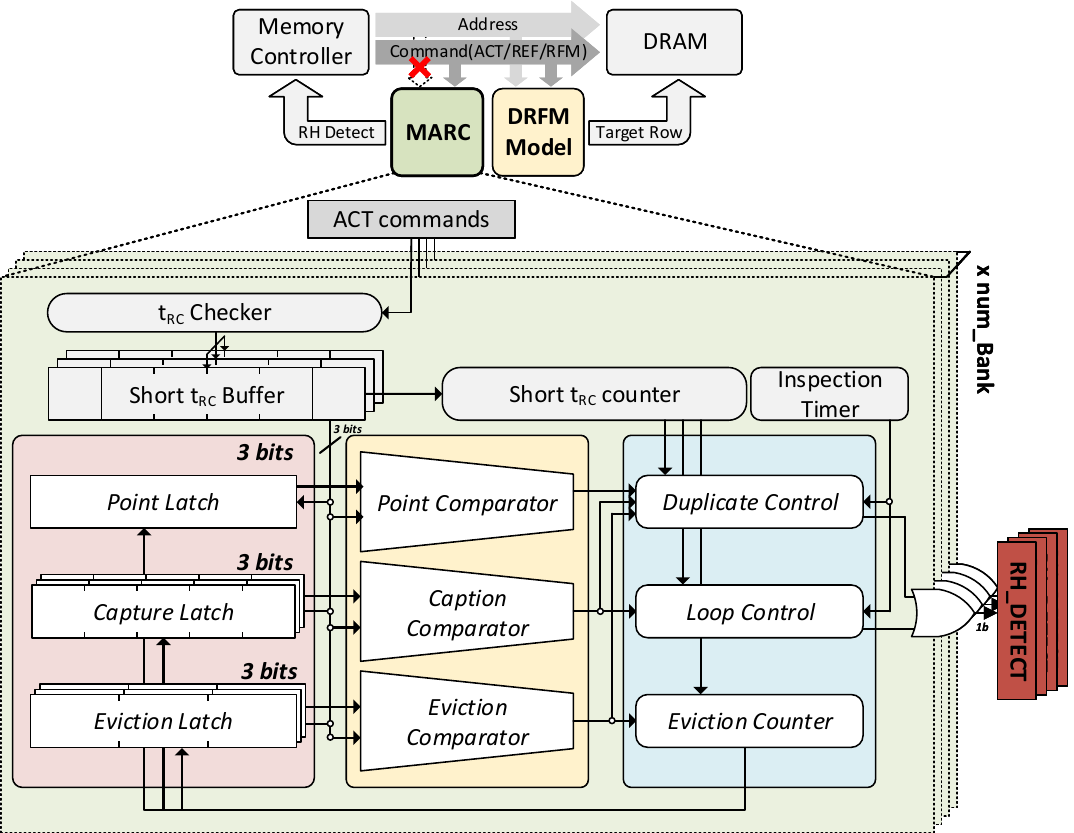}
\vspace{-0.1in}
\caption{Overall Architecture of \sysname}
\label{figure_Architecture}
\end{figure}

\subsection{Defining the Malicious Attack Pattern}
\label{sec_arch_A}
Based on the observation in Section~\ref{observation_sec}, we define the malicious RH attack pattern using the patterns of $t_{RC}$. This approach provides a significant advantage of \sysname, an immunity to the scale-down issue of DRAM. Unlike the counter-based tracking scheme, where its hardware overhead grows exponentially to the scale-down issue (i.e., reduction of cell endurance and the increasing number of rows), \sysname eliminates the need to track aggressors and instead uses $t_{RC}$ to detect RH attack patterns. It provides a maintaining hardware overhead from the scale-down issue. As in the normal workload, the majority of $t_{RC}$ is huge, and only less than 1\% is \textit{short $t_{RC}$}. However, due to the abnormal use of DRAM, the majority of $t_{RC}$ appears to be short in the RH attack pattern. This distinct characteristic allows \sysname to distinguish the RH attack pattern and the normal workload. \sysname examines two conditions: 1) whether the number of \textit{short $t_{RC}$} exceeds the \textit{short $t_{RC}$} threshold (\textit{$S\_t_{RC\_TH}$}), and 2) whether there is a sequence of repetitive pattern.

We introduce \textit{short $t_{RC}$} threshold, \textit{$S\_t_{RC\_TH}$}, the minimum number of \textit{short $t_{RC}$} allowed within the $t_{REFi}$ window interval. This is the minimum requirement for the RH attack pattern. If the number of \textit{short $t_{RC}$} exceeds the \textit{$S\_t_{RC\_TH}$} in a $t_{REFi}$, it is considered as s RH attack pattern. It is crucial to emphasize the frequency of \textit{short $t_{RC}$} values in this context since the common characteristics observed in malicious RH attack patterns include a high prevalence of ACT commands with \textit{short $t_{RC}$} values.

In addition, we define two types of sequence patterns: duplication and looping. The duplication pattern is when identical $t_{RC}$s are repetitively occurred more than two $t_{REFi}$ time periods. The looping pattern is defined when identical sequences of $t_{RC}$s repetitively occur more than two $t_{REFi}$ time periods. Two $t_{REFi}$ time periods are our set minimum threshold, and \sysname continuously monitors whether duplication or looping patterns are maintained across the entire pattern. There are two reasons why \sysname looks for these patterns. First, malicious RH attacks have a clear intention to compromise the integrity of DRAM and induce bit-flips in DRAM cells to access unauthorized data. Malicious RH attacks craft patterns by establishing a consistent $t_{RC}$, expanding aggressor behaviors, and identifying conditions that allow them to circumvent RH mitigation IP defenses. Second, it is worth noting that relying solely on a single value as a criterion is not highly effective due to the multitude of evasion methods and the need to account for system variations during measurement. Hence, we have defined the second criterion as the presence of patterns composed of looping or duplicated sequences of $t_{RC}$ values in the overall pattern of malicious RH attack patterns.

\subsection{Overall Architecture}
Figure~\ref{figure_Architecture} illustrates the architecture of \sysname. It is well-suited to occupy the location of the ARFM model. Thanks to the characteristics of ARFM, the first advantage of \sysname is its ability to detect malicious RH attack patterns independently of the address, allowing us to eliminate unnecessary routing and loading on the address bus. With \sysname integrated into the ARFM model, we can confirm that the connection of the address bus, as depicted in the figure, is no longer necessary.

\sysname is composed of a $t_{RC}$ checker, \textit{short $t_{RC}$} counter, \textit{short $t_{RC}$} buffer, a point latch, capture latches, eviction latches, and control circuits. The $t_{RC}$ checker measures the $t_{RC}$ time between two consecutive ACT commands. As mentioned in Section~\ref{observation_sec}, we define the \textit{short $t_{RC}$} as time period below 100${ns}$, so the range of $t_{RC}$ managed by \sysname is 60${ns}$ ($t_{RCmin}$) $<=$ \textit{short $t_{RC}$} $<=$ 100${ns}$. Instead of using precise $t_{RC}$ values, \sysname exclusively relies on encoded $t_{RC}$ labels to detect malicious RH attack patterns. It encodes them into short-A, short-B, short-C, short-D, and L-$t_{RC}$ with a resolution of 10${ns}$ intervals. The encoded $t_{RC}$ values are stored in the \textit{short $t_{RC}$} buffer, where the number of entries is calculated by dividing $t_{REFi}$ by $t_{RCmin}$.
Not only is \sysname immune to reductions in cell endurance, but the use of this encoding technique also makes it a lightweight hardware solution with minimal area overhead. The data size is reduced to 3 bits, which significantly reduces hardware overhead. It means that the size of the \textit{short $t_{RC}$} buffer's entry, comparators, and latches are only 3 bits. 

The point latch stores a \textit{short $t_{RC}$}, which is a candidate $t_{RC}$ for the duplication attack pattern when the identical $t_{RC}$ are repetitively offloaded to the memory. The capture latch stores \textit{K} number of \textit{short $t_{RC}$} which are candidate $t_{RC}$s for the looping attack pattern when there is a pattern in the $t_{RC}$. The eviction latch stores \textit{k}-2 number of $t_{RC}$, which are different from the $t_{RC}$ in the point latch or the previous pattern in the capture latch. The control circuit manages the detection of the RH attack pattern in duplication and loop conditions, as well as the reset control.

\begin{figure}[!t]
\centering
\includegraphics[width=\columnwidth]{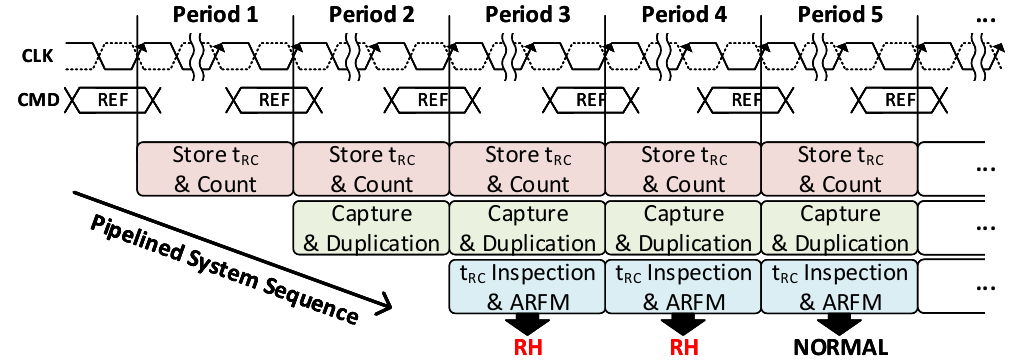}
\vspace{-0.25in}
\caption{\sysname System Operation Flow}
\vspace{-0.2in}
\label{figure_System_Sequence}
\end{figure}

\subsection{Operation Flow}

The \sysname detects real-time malicious RH attack patterns in a pipeline as shown in Figure~\ref{figure_System_Sequence}. It includes three phases: storing \& counting $t_{RC}$, capturing \& duplicating, and inspecting $t_{RC}$ \& activating ARFM. Each phase processes data within one $t_{REFi}$ window. Initially, \sysname tallies and stores \textit{short $t_{RC}$}s in a buffer. It then examines these values, selecting the most frequent pattern for further analysis. If a repeating pattern is identified over a set period, \sysname alerts the memory controller to activate ARFM at the initial level-A RAAIMT. Throughout ARFM's activation, \sysname continuously checks for pattern repeats, adjusting ARFM's intensity based on the repetition duration. If a pattern matches our defined RH attack criteria, \sysname instructs the MC to escalate the response, progressing through levels until reaching level-C.

In the capture \& duplication phase, detailed in Figure~\ref{figure_capture_system} for K=3, the system commands the $t_{RC}$ buffer to identify patterns. This phase involves three latches (point, capture, eviction) and comparators (point, capture, eviction). The point latch stores a single entry, the capture latch stores three entries, and the eviction comparator stores a single entry. Each comparator checks its latch against a \textit{short $t_{RC}$} from the buffer, playing a key role in pattern detection.

The capture system processes each \textit{short $t_{RC}$} in the buffer as follows: In the \textbf{point process} (\textbf{T$_{1}$} to \textbf{T$_{2}$} in Figure~\ref{figure_capture_system}), the first $t_{RC}$ label fills the point latch (e.g.,(\textbf{C} in the latch). Then, the next label is checked against the point latch's value with the point comparator. If a match occurs, the point compare flag activates, continuing the search in the buffer until a different label is found. In the process repetition, if a new label differs from the point latch's, it enters the capture latch's first slot, ending the point and starting the \textbf{capture process} (e.g., \textbf{D} into capture latch, \textbf{T$_{2}$} to \textbf{T$_{6}$} in Figure~\ref{figure_capture_system}). Initially, labels are checked against the capture latch's first value. If matching, the capture flag activates, and the search continues for a new label. This comparison repeats until the capture latch nearly fills (e.g., \textbf{D} and \textbf{C} into the latch, \textbf{T$_{2}$} to \textbf{T$_{5}$}). The last slot filling is unique; a new label fills it, regardless of similarity to the last entry (e.g., \textbf{C} fills the final slot, \textbf{T$_{6}$}). This fills the capture latch completely.

\begin{figure}[!t]
\centering
\includegraphics[width=\columnwidth]{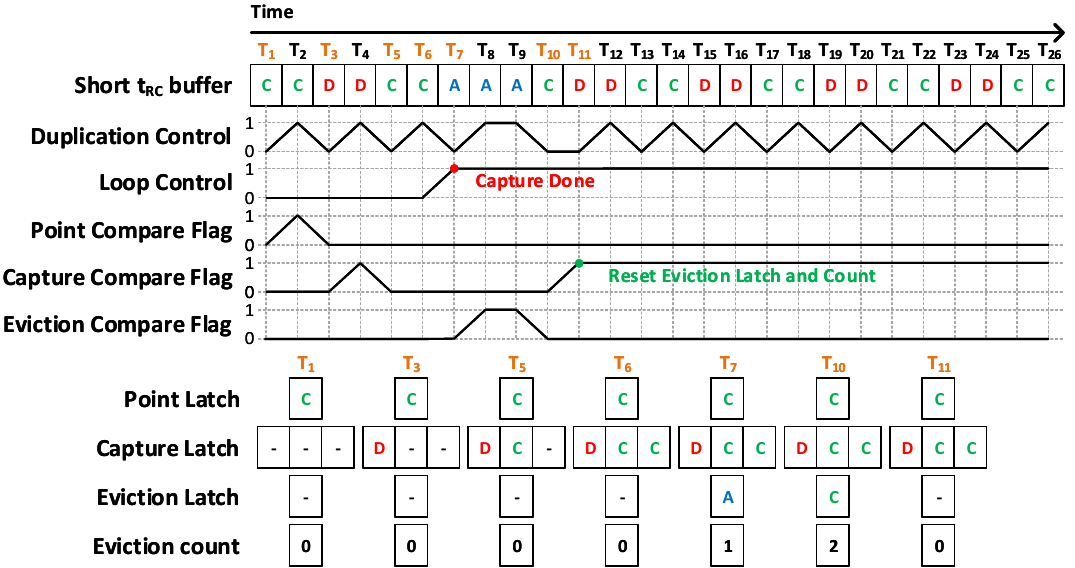}
\vspace{-0.25in}
\caption{Operation Flow of Capture \& Duplication Period }
\vspace{-0.2in}
\label{figure_capture_system}
\end{figure}

This capture system's core function is to monitor whether capture latch values repeat over time. Once the latch fills and detects a new label, the capture ends, triggering a loop control signal ((\textbf{T$_{7}$} in Figure~\ref{figure_capture_system})). The \textbf{eviction process} starts when opposing patterns arise, increasing the eviction count if a new, different label is found (\textbf{A} into eviction latch, \textbf{T$_{7}$}). Matching labels activate the eviction flag (\textbf{T$_{8}$}), enabling the detection of new patterns. Exceeding the eviction threshold indicates a wrong pattern, prompting a reset to the initial process. But, if any compare flags are set, it halts the reset, maintaining the current state (\textbf{T$_{11}$}). If any of the compare flags are activated during the capture \& duplication period, a duplicated control signal is turned on. This signal, along with the loop control signal, becomes an operand in the following inspection \& ARFM phase, where \sysname detects an RH attack. If either of these signals is activated during the whole inspection phase, \sysname recognizes it as an RH attack. Through various recognition systems, \sysname can capture the diverse patterns that occur during an RH attack.

\section{Methodology}
\label{Section_meth}

\begin{table}[!t]
\centering
\caption{Simulation System Configuration}
\vspace{-0.1in}
\includegraphics[width=\columnwidth]{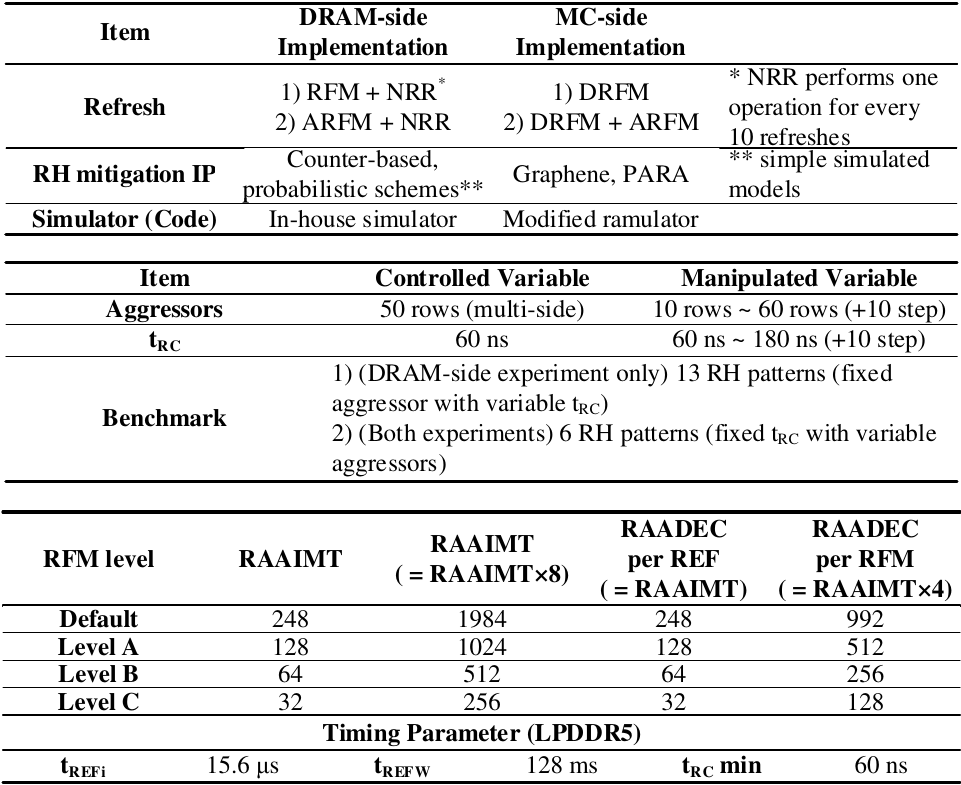}
\vspace{-0.35in}
\label{table_methodology}
\end{table}

\textbf{Hardware Setup.} RH mitigation IP can be integrated into both DRAM and MC, with the victim row curing method varying depending on the integration location. In DRAM, the basic operation involves using the REF command for NRR, and the curing operation can also be performed using the RFM command. Conversely, IPs mounted on the MC-side are unable to perform NRR operations. Only aggressors detected by DRFM can be communicated to DRAM. Owing to these differences in curing methods, our experiments are divided into DRAM-side and MC-side evaluations based on the RH mitigation IP’s integration location.

By conducting separate evaluations for both the DRAM-side and MC-side, we aim to demonstrate that ARFM support by \sysname enhances defense capability, irrespective of the RH mitigation IP's location. Table~\ref{table_methodology} presents the hardware setup conditions for our experiment.

In the DRAM-side section, we utilize the simple probabilistic scheme and counter-based tracking scheme, as introduced in Section~\ref{observation_sec}, to assess the trends in each mechanism. We have designed the \sysname module to integrate with a simple simulated model and evaluate it through an in-house simulator, ensuring it aligns with DRAM specifications. 

On the MC-Side, we employ Graphene and PARA, which are exemplary IPs representing the counter-based tracking and probabilistic mechanisms, two leading methods in the domain of open-source RH mitigation mechanisms. Utilizing these validated IPs as our RH mitigation IPs bolsters the credibility of our experimental results. For the system performance evaluation, we have devised Ramulator~\cite{kim2015ramulator, yauglikcci2021blockhammer} by integrating the \sysname module with each of these two IPs and configuring it to support RFM operation. To estimate DRAM energy consumption, we use DRAMPower. We model the area and energy overhead of \sysname's hardware by implementing RTL design and synthesizing it using Synopsys Design Compiler with Samsung 28nm standard cell library. 

\textbf{Benchmark.} 
The strength of RH defenses depends on various factors, including the configuration of the attack pattern (i.e., single-sided, double-sided, or multi-sided approaches), the number of aggressors, and the design of the RH mitigation IP. Considering the variability of these factors, it is not feasible to obtain absolute metrics for defense capability. 

As a benchmark, our experimental design included two key scenarios: first, an experiment with a fixed aggressor count of 50, employing a multi-sided approach, where the $t_{RC}$ value was varied in 10${ns}$ increments; and second, an experiment with a fixed $t_{RC}$ of 60${ns}$, where the number of aggressors was altered in increments of 10. These configurations resulted in a comprehensive array of 19 different patterns, from which we derived 100 distinct MER values. Detailed descriptions and specifics of our benchmark are collated in Table~\ref{table_methodology}.

\textbf{DRAM Parameter.} In this experiment, setting the RAAIMT value is crucial, and our criteria for this setting are detailed in Table~\ref{table_methodology}. In consideration of RFMTH, we have chosen a default value of 248. This value provides approximately a 10\% margin over the ratio of $t_{REFi}$ to $t_{RCmin}$. For the ARFM operation, we adjusted the RFM level to be reduced by 50\% at each stage. The RAAMMT was set to 8 $\times$ RAAIMT, and the RAADEC was set to 4 $\times$ RAAIMT. This configuration corresponds to the scenario with the least RFM usage, thereby minimizing power consumption under normal workloads. Other timing parameters were fixed based on LPDDR5 specifications, with $t_{REFw}$, $t_{REFi}$, and $t_{RCmin}$  set at 128${ms}$, 15.6${us}$, and 60${ns}$, respectively. 
\section{Evaluation}
\label{sec_eval}
In this section, we examine the performance of RH mitigation IPs when combined with \sysname. In Section~\ref{sec_eval_a}, we first investigate the hardware specifications and performance of \sysname. Then, in Section~\ref{sec_eval_b}, we observe the max exposure of RH mitigation IPs on the DRAM-side with and without the integration of \sysname. Script-based evaluations on the DRAM-side offer advantages in terms of pattern scalability, allowing us to generate a sufficient number of RH attack patterns and secure results on max exposure improvements due to \sysname for each pattern. As representative results of this extensive experimentation, we present the MER based on the number of aggressors and various $t_{RC}$. In Section~\ref{subsec_eval_C}, we observe the max exposure of RH mitigation IPs on the MC-side with and without the integration of \sysname. We provide MER based on the number of aggressors and examine the power increase caused by \sysname in RH attack scenarios.

\subsection{\sysname Hardware Module}
\label{sec_eval_a}
\textbf{Area and Power.} \sysname is a compact hardware module to defend against RH attack and is even more effective to the DRAM's scale-down issue. Unlike the previous counter-based tracking schemes (e.g., Graphene and TWiCE), the number of entries in the short $t_{RC}$ buffer remains fixed, regardless of the increasing number of DRAM rows or the decrease in cell endurance. Even though the short $t_{RC}$ buffer is the only major area overhead of \sysname, it only has 260 entries, and each entry has the size of 3 bits. Thanks to the 3-bit data type of encoded $t_{RC}$ label, the comparators are also light. According to our synthesis result, \sysname's hardware area is 3363 \textit{$\mu m^2$}, which is 23\% smaller than that of Graphene, and power usage is 1.3 $mW$ per bank.






\begin{table}[!t]
\centering
\caption{\sysname Detection Efficacy on Various $t_{RC}$ Patterns}
\vspace{-0.1in}
\includegraphics[width=\columnwidth]{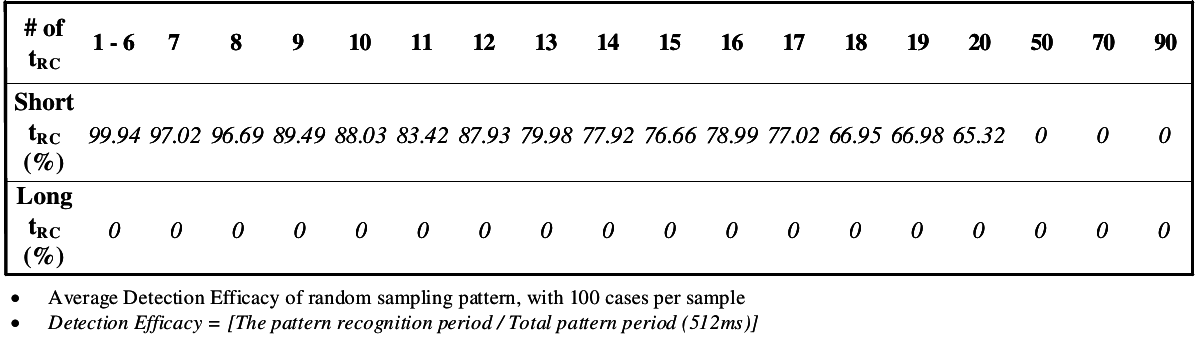}
\label{table_eval}
\end{table}

\textbf{Performance.} As mentioned in Section~\ref{sec_arch_A},  \sysname examines two conditions of malicious RH attack pattern: 1) whether the number of short $t_{RC}$ exceeds \textit{$S\_t_{RC\_TH}$}, and 2) whether there is a sequence of repetitive pattern. Based on these conditions, we construct the benchmark with various $t_{RC}$ patterns. We construct $t_{RC}$ combinations and expand the number of $t_{RC}$ elements in each pattern list, ranging from 1 (duplication pattern) to 100 or fewer (looping pattern). For instance, if the number of $t_{RC}$ = 1 in the short $t_{RC}$ combination pattern, the pattern is [65 65 65 65 ...], and for the number of $t_{RC}$=3, the $t_{RC}$ pattern is formed as [65 93 100 65 93 100 ...]. It is likely to be an RH attack pattern with short $t_{RC}$ with the less number of $t_{RC}$ value. In contrast, it is likely to be a normal workload pattern with long $t_{RC}$ with more $t_{RC}$ values. For the experiment, the size of the capture latch, \textit{K}, is set to 3 to detect extreme RH attack patterns exclusively. If we increase \textit{K}, \sysname tends to detect relatively less extreme RH attack patterns.



Table~\ref{table_eval} shows the outcomes of \sysname's detection efficacy across a spectrum of ${t_{RC}}$ combination patterns. In our assessment of \sysname's performance, we categorized short ${t_{RC}}$s, defined as 100ns or shorter, into 23 distinct pattern cases (ranging from 1 to 20, and including 50, 70, 90 each) based on the quantity of ${t_{RC}}$s. For each case, 100 random tRC combination patterns were generated. The rate of RH attack pattern recognition by \sysname is quantified by the proportion of the pattern recognition duration over the total pattern duration. As elucidated in Table~\ref{table_eval}, \sysname exhibits a recognition rate exceeding 99\% for rudimentary combination patterns involving short ${t_{RC}}$s. Furthermore, the identification of patterns that \sysname does not recognize corroborates its accurate functionality and the designed operational behavior, underscoring the system's efficacy in distinguishing between varying patterns and its precision in executing its intended defensive strategy against RH attacks.

Patterns not recognized by \sysname exhibit two primary trends. The first involves patterns consisting of long $t_{RC}$s with values exceeding 100${ns}$. In these cases, \sysname does not classify them as malicious RH attack patterns due to the non-fulfillment of the first pattern recognition condition. In the case of patterns where $t_{RC}$ exceeds 100${ns}$, it has already been confirmed in Figure~\ref{fig_MER_NRR_para_gra} that as ${t_{RC}}$ increases, the severity of RH attack is greatly weakened. 

The second trend occurs when there are more than 20 combinations of $t_{RC}$. Here, \sysname determines that the second pattern recognition condition is not met. Notably, \sysname tracks $t_{RC}$ values based on encoded levels, meaning that for \sysname to classify a pattern as non-malicious according to the second condition, the deviation in $t_{RC}$ values must be significantly large, typical of normal workloads, and thus not recognized. However, if a pattern has more than 20 different $t_{RC}$ values, \sysname will still recognize it as significant if the encoding levels are identical. The result highlights \sysname's capability to discern between malicious RH attack patterns and normal workloads efficiently, ensuring power efficiency and effectiveness in RH attack detection.

\begin{figure}[!t]
\centering
\includegraphics[width=0.95\columnwidth]{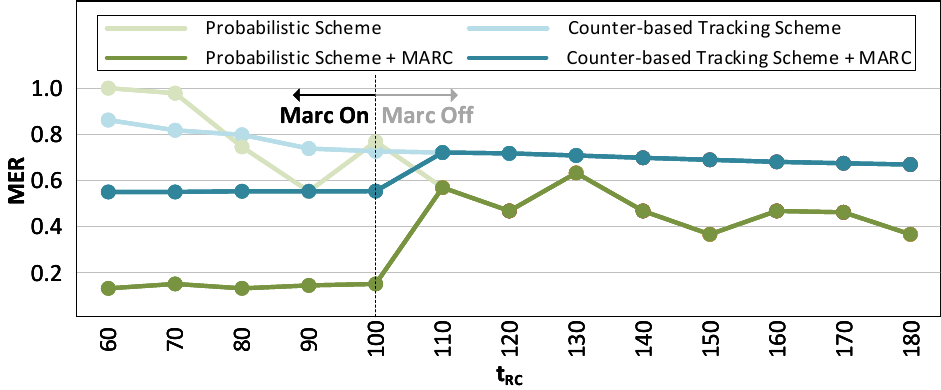}
\caption{Max Exposure Ratio of DRAM-side \sysname on Various $t_{RC}$}
\vspace{-0.25in}
\label{figure_eval_dramside_tRC}
\end{figure}

\subsection{DRAM-Side Evaluation}
\label{sec_eval_b}
Our objective is to enhance defense against RH attacks characterized by short $t_{RC}$ values while relying on system-level support without requiring any hardware changes. In this study, we defined short $t_{RC}$ as 100${ns}$. We conduct a set of experiments comparing the two vanilla RH mitigation IPs with the integration of RH mitigation IP with \sysname. They are denoted as the probabilistic scheme, counter-based tracking scheme, probabilistic scheme + \sysname, and counter-based tracking scheme + \sysname. Figure~\ref{figure_eval_dramside_tRC} illustrates the result of MER of each scheme. The result shows that the probabilistic scheme + \sysname and counter-based tracking scheme + \sysname within the short $t_{RC}$ area significantly enhances defense performance through the utilization of ARFM.

The counter-based tracking scheme + \sysname illustrates that even short $t_{RC}$ attack patterns converge to the internal tracking logic threshold level, resulting in a 1.5$\times$ improvement in MER compared to the vanilla counter-based tracking scheme. In contrast, the probabilistic scheme + \sysname exhibits an impressive 8.1$\times$ improvement in MER compared to the vanilla probabilistic scheme. This substantial improvement is attributed to the probabilistic scheme's approach of refreshing rows in proportion to the number of added RFM commands, thereby increasing the likelihood of refreshing victim rows and enhancing random sampling probability as the RFM period shortens. On the other hand, the counter-based tracking scheme relies on a logic threshold for determining whether the internally tracked row is an aggressor. Consequently, even with an increased number of RFM commands due to ARFM, it refrains from refreshing until it identifies itself as an aggressor, resulting in potential RFM wastage.

For a more detailed confirmation of the improvement effect, we present the results of a max exposure experiment while varying the number of aggressors, focusing on the most critical $t_{RCmin}$ (= 60${ns}$) attack level, as shown in Figure~\ref{figure_eval_dramside_aggressor}. While the max exposure of the probabilistic scheme may vary depending on the attack pattern, the probabilistic scheme + \sysname consistently exhibits a 10$\times$ improvement than the vanilla probabilistic scheme. In addition, the counter-based tracking scheme demonstrates nearly uniform max exposure for aggressor patterns smaller than the table size, and even with an increased number of RFM commands, it converges to the internally determined logic threshold. The counter-based tracking scheme + \sysname improves its MER by 3.1$\times$ than the vanilla counter-based tracking scheme. Although there may be variations in the degree of improvement, it is evident that max exposure improves in the worst-case scenarios for both IP schemes with the assistance of ARFM.

\begin{figure}[!t]
\centering
\includegraphics[width=0.95\columnwidth]{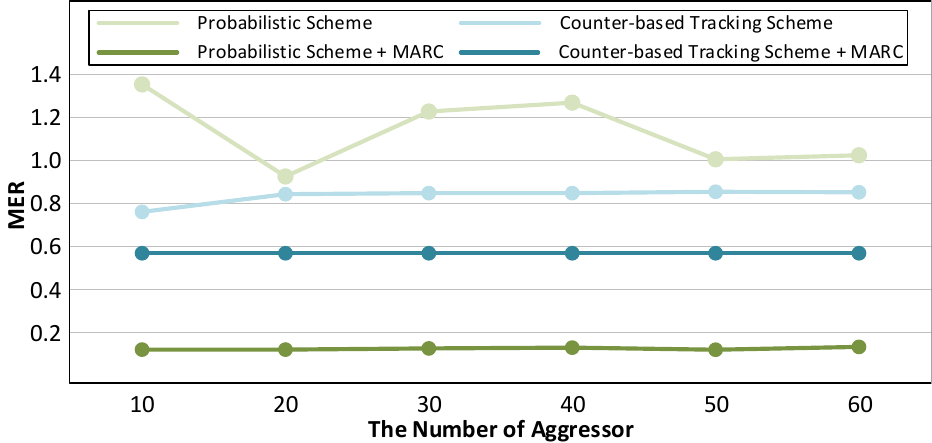} 
\caption{Max Exposure Ratio of DRAM-side \sysname for Various Number of Aggressors}
\vspace{-0.15in}
\label{figure_eval_dramside_aggressor}
\end{figure}

\subsection{MC-Side Evaluation}
\label{subsec_eval_C}
Experiments are conducted while expanding the aggressor on PARA, PARA + \sysname, Graphene, and Graphene + \sysname when integrated into the MC-side. The results show a similar trend to the DRAM-side evaluation results, as shown in Figure~\ref{figure_eval_MCside_aggressor}. As is well-known, under conditions where \sysname is not integrated, Graphene exhibits superior defense capabilities compared to PARA. However, when \sysname is integrated under these conditions, PARA's defense ability improves by more than 50$\times$, surpassing the performance of Graphene. This trend can be attributed to the convergence of the logic threshold determined by Graphene's internal process. Additionally, it's worth noting that the max exposure of Graphene has also improved by more than 5.7$\times$ through the integration of \sysname.

\begin{figure}[!t]
\centering
\includegraphics[width=0.95\columnwidth]{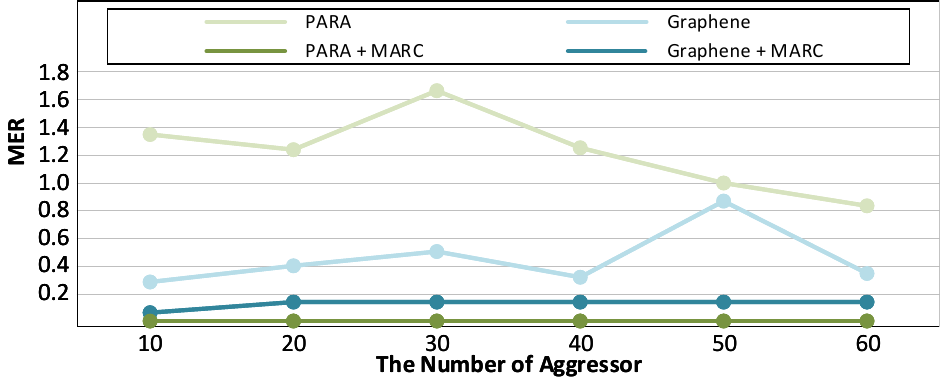}
\caption{Max Exposure Ratio of Memory Controller-side \sysname for Various Number of Aggressors}
\vspace{-0.2in}
\label{figure_eval_MCside_aggressor}
\end{figure}

\subsection{Power Consumption}
\label{subsec_eval_D}
In reviewing Figure~\ref{figure_tRCNWPattern} and \ref{figure_shortTrc}, we confirmed that the frequency of short $t_{RC}$s under 100${ns}$ in normal workloads is exceedingly low, and Table~\ref{table_eval} indicates that \sysname remains inactive for $t_{RC}$ combination patterns above 100${ns}$. This leads to the conclusion that the presence of \sysname does not increase power consumption in normal workloads. Figure~\ref{figure_eval_MCside_power}, which depicts power consumption in RH attack patterns where \sysname is active, counters the widespread belief that power consumption escalates with the additional issuance of RFM. The operational pattern under which \sysname functions is characterized by a high incidence of ACT commands within short $t_{RC}$s. Consequently, the marginal increase in power consumption due to \sysname's extra RFM commands is relatively insignificant when compared to the overall power usage by ACT commands. Hence, as \sysname does not cause a noticeable rise in power consumption in RH attack scenarios, it can be deduced that \sysname does not contribute to power consumption concerns in either normal workloads or RH attack patterns.
 
\section{Related Work \& Discussion}
\label{sec_discussion}

\textbf{ARR.} PARA, CBT, TWiCe, and Graphene~\cite{kim2014flipping, seyedzadeh2016counter, lee2019twice, park2020graphene} are designed based on adjacent row refresh (ARR), which reactively issue an ACT command for curing a victim row. It can send an additional command immediately. However, ARR cannot be applied in practice since it requires a major modification of a passive DRAM.

\textbf{RFM.} Mithril~\cite{kim2022mithril} is designed based on RFM, which gives an opportunity for both DRAM and MC to cooperate to defend against RH attacks. With the support of RFM in a JEDEC standard, DRAM and MC have their own way of defending RH attacks. 

\textbf{Workload Analysis.} 
In this study, we selected a mobile product as the representative model for workload analysis, focusing on the refresh interval ($t_{REFi}$) as a key criterion. DDR5 and GDDR7 have refresh intervals of 3.9${us}$ and 1.9${us}$, respectively, which are significantly shorter than LPDDR5's 15.6${us}$ interval at cold temperature. This discrepancy in $t_{REFi}$ impacts the number of refresh commands a DRAM can issue during the same period, influencing its ability to respond to RH attacks effectively.
Devices with shorter refresh intervals can issue more commands in the same timeframe, offering an advantage against RH attacks. Mobile devices, with typically longer $t_{REFi}$, are at a disadvantage. Enhancing RH defense in mobile devices could improve resilience across all products. If \sysname improves maximum exposure in mobile settings with fewer refresh commands, it will likely have an even greater effect in other contexts with more frequent refreshes. This suggests \sysname's potential to significantly bolster RH defense across various environments.

\textbf{Discussion.} 
Our study shows that using ARFM at the system level significantly enhances RH defense without additional hardware. The probabilistic scheme benefits greatly from more RFM commands, unlike the costlier counter-based tracking. This finding challenges the belief that counter-based tracking is inherently more effective for RH mitigation. The probabilistic method effectively uses increased RFM frequency to boost curing chances, while the counter-based scheme does not. RH defense effectiveness depends on both hardware and the frequency of curing affected addresses.

Although counter-based tracking is effective against RH attacks, it needs refinement with substantial system support. To maximize benefits in such environments, developing and integrating complementary strategies is essential. This need has driven research focused on optimizing ARFM support. As this research progresses, we expect a more robust RH defense system, especially with proactive system support.

\begin{figure}[!t]
\centering
\includegraphics[width=\columnwidth]{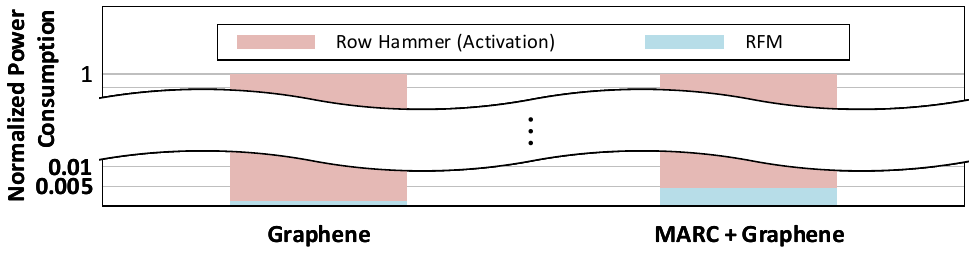}
\caption{Normalized Power Consumption of \sysname}
\vspace{-0.2in}
\label{figure_eval_MCside_power}
\end{figure}
 
\section{Conclusion}
Establishing a hardware-centric RH defense system faces significant limitations, including increased power demands and size constraints, as the cell endurance continues to degrade. To address these challenges, a paradigm shift is necessary: moving from a hardware-centric to a system-level support-based RH defense system utilizing RFM. This paradigm shift entails two critical factors: 1) implementing strategies for the efficient use of RFM, and 2) developing RH mitigation IPs that incorporate RFM considerations. To this end, our study focuses on the efficient utilization of RFM, addressing one of the key factors. We propose a method of adaptive MC support for malicious RH attack patterns using ARFM, where \sysname identifies these patterns by analyzing the $t_{RC}$ value and pattern. \sysname is a compact and smart module that is 23\% smaller than Graphene and has no size overhead due to deterioration of cell endurance. 
Our experiments demonstrate that the probabilistic scheme effectively aligns with RFM, whereas the counter-based tracking scheme shows limited effectiveness in RFM support. Consequently, future research in RH mitigation IP should focus on RFM integration, the second critical factor of this paradigm shift. The development of such IPs will optimize the effectiveness of \sysname in RH defense systems.



\end{document}